\documentclass[aip, pop, reprint]{revtex4-2}

\usepackage{amssymb}
\usepackage{amsmath}
\usepackage{float}
\usepackage{graphicx}
\usepackage[normalem]{ulem}

\newcommand{\uwmadison}{University of Wisconsin--Madison, Madison, Wisconsin, USA}
\newcommand{\llnl}{Lawrence Livermore National Laboratory, Livermore, California, USA}
\newcommand{\ipp}{Max-Planck-Institute f\"ur Plasmaphysik, Garching, Germany}
\newcommand{\theaenergy}{Thea Energy, Kearny, New Jersey, USA}
\newcommand{\helicityspace}{HelicitySpace, Pasadena, California, USA}
\newcommand{\kit}{Kyoto Institute of Technology, Kyoto, Japan}
\newcommand{\atsrt}{54.7}
\newcommand{\tpmatsrt}{125.3}
\newcommand{\lp}{\left(}
\newcommand{\rp}{\right)}
\newcommand{\la}{\langle}
\newcommand{\ra}{\rangle}

\newcommand{\ls}{\left[}
\newcommand{\rs}{\right]}
\newcommand{\lc}{\left\{}
\newcommand{\rc}{\right\}}
\newcommand{\fr}{\frac}
\newcommand{\T}{\text}
\newcommand{\bs}{\boldsymbol}
\newcommand{\pd}{\partial}
\newcommand{\curl}{\boldsymbol{\nabla}\times}
\newcommand{\whbs}[1]{\widehat{\boldsymbol{#1}}}
\newcommand{\x}{\times}
\newcommand{\cd}{\cdot}

\newcommand{\etc}{\textit{etc}}
\newcommand{\eg}{\textit{e}.\textit{g}}
\newcommand{\ie}{\textit{i}.\textit{e}}
\newcommand{\degs}{^\circ}
\newcommand{\phia}{\phi_\text{a}}
\newcommand{\phib}{\phi_\text{b}}
\newcommand{\phic}{\phi_\text{c}}
\newcommand{\phid}{\phi_\text{d}}
\newcommand{\phie}{\phi_\text{e}}
\newcommand{\phif}{\phi_\text{f}}
\newcommand{\thea}{\theta_\text{a}}
\newcommand{\theb}{\theta_\text{b}}
\newcommand{\thec}{\theta_\text{c}}
\newcommand{\thed}{\theta_\text{d}}
\newcommand{\thee}{\theta_\text{e}}
\newcommand{\thef}{\theta_\text{f}}

\newcommand{\Ia}{I_\text{a}}
\newcommand{\Ib}{I_\text{b}}
\newcommand{\Ic}{I_\text{c}}
\newcommand{\Id}{I_\text{d}}
\newcommand{\Ie}{I_\text{e}}
\newcommand{\If}{I_\text{f}}
\newcommand{\xa}{x_\text{a}}
\newcommand{\ya}{y_\text{a}}
\newcommand{\za}{z_\text{a}}
\newcommand{\xb}{x_\text{b}}

\newcommand{\bxa}{\bar{x}_\text{a}}
\newcommand{\bya}{\bar{y}_\text{a}}
\newcommand{\bza}{\bar{z}_\text{a}}
\newcommand{\bxb}{\bar{x}_\text{b}}
\newcommand{\byb}{\bar{y}_\text{b}}
\newcommand{\bzb}{\bar{z}_\text{b}}
\newcommand{\bxc}{\bar{x}_\text{c}}
\newcommand{\byc}{\bar{y}_\text{c}}
\newcommand{\bzc}{\bar{z}_\text{c}}
\newcommand{\bxd}{\bar{x}_\text{d}}
\newcommand{\bvx}{\bar{v}_x}
\newcommand{\bvy}{\bar{v}_y}
\newcommand{\bvz}{\bar{v}_z}
\newcommand{\bxi}{\bar{x}_i}
\newcommand{\byi}{\bar{y}_i}
\newcommand{\bzi}{\bar{z}_i}
\newcommand{\bxj}{\bar{x}_j}
\newcommand{\byj}{\bar{y}_j}
\newcommand{\bzj}{\bar{z}_j}
\newcommand{\bxk}{\bar{x}_k}
\newcommand{\byk}{\bar{y}_k}
\newcommand{\bzk}{\bar{z}_k}
\newcommand{\bxl}{\bar{x}_l}
\newcommand{\byl}{\bar{y}_l}
\newcommand{\bzl}{\bar{z}_l}
\newcommand{\bxm}{\bar{x}_m}
\newcommand{\bym}{\bar{y}_m}
\newcommand{\bzm}{\bar{z}_m}
\newcommand{\xij}{x_{ij}}
\newcommand{\yij}{y_{ij}}
\newcommand{\zij}{z_{ij}}
\newcommand{\xad}{x_\text{ad}}
\newcommand{\yad}{y_\text{ad}}
\newcommand{\zad}{z_\text{ad}}
\newcommand{\xbe}{x_\text{be}}
\newcommand{\ybe}{y_\text{be}}
\newcommand{\zbe}{z_\text{be}}
\newcommand{\xcf}{x_\text{cf}}
\newcommand{\ycf}{y_\text{cf}}
\newcommand{\zcf}{z_\text{cf}}
\newcommand{\xk}{x_\text{k}}
\newcommand{\yk}{y_\text{k}}
\newcommand{\zk}{z_\text{k}}
\newcommand{\xl}{x_\text{l}}
\newcommand{\yl}{y_\text{l}}
\newcommand{\zl}{z_\text{l}}
\newcommand{\xm}{x_\text{m}}
\newcommand{\ym}{y_\text{m}}
\newcommand{\zm}{z_\text{m}}
\newcommand{\xabcd}{x_\text{abcd}}
\newcommand{\yabcd}{y_\text{abcd}}
\newcommand{\zabcd}{z_\text{abcd}}
\newcommand{\xbcda}{x_\text{bcda}}
\newcommand{\ybcda}{y_\text{bcda}}
\newcommand{\zbcda}{z_\text{bcda}}
\newcommand{\xacbd}{x_\text{acbd}}
\newcommand{\yacbd}{y_\text{acbd}}
\newcommand{\zacbd}{z_\text{acbd}}
\newcommand{\Ip}{I_\text{p}}
\newcommand{\Bt}{B_\text{t}}
\newcommand{\nel}{n_\text{e}}
\newcommand{\Te}{T_\text{e}}

\begin{document}

\title{Three-dimensional unmagnetized Mach probe analysis and initial flow measurements in reversed-field pinch experiments}
\author{K.~J.~McCollam}\email{kmccollam@wisc.edu.}
\affiliation{\uwmadison}
\author{R.~Reksoatmodjo}
\affiliation{\llnl}
\author{J.~von~der~Linden}
\affiliation{\ipp\ (Current address: \theaenergy)}
\author{J.~Sears}
\affiliation{\llnl}
\author{S.~You}
\affiliation{\helicityspace}
\author{H.~Himura}
\affiliation{\kit}
\author{A.~F.~Almagri}
\affiliation{\uwmadison}
\author{M.~Reyfman}
\affiliation{\uwmadison}
\author{C.~C.~Rouda}
\affiliation{\uwmadison}
\author{J.~S.~Sarff}
\affiliation{\uwmadison}
\author{A.~M.~Sellner}
\affiliation{\uwmadison}

\date{\today}

\begin{abstract}
A novel matrix method of analyzing ion saturation current data from a general three-dimensional (3D) array of unmagnetized Mach probe tips is developed and used with data sets from two 3D Mach probes to make initial measurements of local plasma flow velocity in reversed-field pinch (RFP) experiments in the Madison Symmetric Torus (MST).
The two 3D Mach probes are composed of regular polyhedral arrays of six and four tips, respectively, with the six-tip array composed of three orthogonal pairs of mutually opposite tips at the vertices of a regular octahedron and the four-tip array composed of non-opposite tips at the vertices of a regular tetrahedron, the analysis of which is specifically facilitated by the matrix method.
Velocity measurement uncertainties for the Mach probes are derived based on uncertainties in probe machining and ion saturation current measurements, and typical relative uncertainties for the probes are estimated to be of order several percent, likely smaller than systematic uncertainties related to the Mach probe calibration constant and experimental uncertainties related to plasma and probe conditioning.
Initial results for the octahedral probe show flow speeds of roughly the expected magnitudes based on previous MST measurements but with somes differences in flow direction, while those for the tetrahedron probe show similar flow directions to some previous measurements but also some larger than expected speeds.
We consider possible causes for the unexpected results of these initial tests, with a focus on probe conditioning and fast electron issues.
\end{abstract}

\maketitle

\section{Introduction}
\label{sec:intro}

The reversed-field pinch (RFP)
\cite{marrelli_2021}
plasma configuration exhibits plasma flows that can help to reveal underlying transport and relaxation mechanisms influencing its dynamics.
The RFP is a toroidal magnetic confinement system whose name comes from the toroidal reversal of its highly sheared equilibrium magnetic field in the plasma edge region.

As with other toroidal plasmas such as tokamaks or stellarators
\cite{ida_2014},
a prominent example of flow in RFPs is intrinsic or spontaneous equilibrium plasma rotation
\cite{carraro_1998, craig_2019}.
In the RFP, such rotation typically arises in both the toroidal and poloidal directions and is chiefly ascribed to E-cross-B drift in the region from the core to the near edge, with an outward ambipolar radial electric field $E_\T{r}$ generated by relatively rapid electron radial transport compared to ion transport in the presence of a stochastic magnetic field
\cite{ji_1991, tsui_1991},
while in the far edge, a reversed, inward $E_\T{r}$ is sometimes observed
\cite{vianello_2002}
and has been hypothetically ascribed to edge ion recycling
\cite{antoni_1996a},
finite ion Larmor radius
\cite{antoni_1997b},
or a locally non-stochastic magnetic field
\cite{antoni_1997b}.

Such spontaneous equilibrium flows have been measured in various RFP experiments
\cite{den_hartog_1995b, antoni_1996b, carraro_1998, den_hartog_1998, sakakita_1999, sakakita_2000, fontana_1999, malmberg_2004, cecconello_2006, kuritsyn_2009, de_masi_2010, miller_2011, ding_2013}
and have generally been found to show overall consistency in their observed rotation rates and directions with E-cross-B and ion diamagnetic drifts as well as with the mode rotation of magnetic fluctuations
\cite{den_hartog_1995b, den_hartog_1998, malmberg_2004}.
In the Madison Symmetric Torus (MST) experiment
\cite{dexter_1991},
such measurements in the RFP edge
\cite{fontana_1999, miller_2011}
have tended to show equilibrium flows of $\sim20$~km/s in the E-cross-B direction, i.e.\ toroidally in the same sense as the plasma current $\Ip$,
typically with a somewhat smaller poloidal component, while the radial component was observed to be much smaller than either the toroidal or poloidal component.

In addition to such mainly two-dimensional (2D) equilibrium flows, fully three-dimensional (3D) plasma flow fluctuations play key roles in multiple RFP relaxation mechanisms.
In particular, flow fluctuations contributing to the MHD `dynamo' mean-field electromotive force that acts to maintain the RFP equilibrium magnetic structure including the toroidal field reversal during magnetic relaxation
\cite{schnack_1985}
have been directly measured
\cite{den_hartog_1999, fontana_2000, ennis_2010, craig_2017},
as have those contributing to the velocity fluctuation-induced mean-field Reynolds stress involved in momentum relaxation
\cite{kuritsyn_2009}.

One relatively recently identified motivation for local, 3D plasma flow measurements in the RFP is the suggestion of evaluating canonical helicity evolution during magnetic relaxation
\cite{you_2012}.
Canonical helicity is a generalization of the magnetic helicity concept involving both the electron or magnetic helicity and the ion or mass fluid helicity
\cite{you_2012, you_2014, you_2016, von_der_linden_2018},
and one key ingredient is the fluid helicity density proportional to $\bs{v}\cd\curl\bs{v}$, i.e.\ the scalar product of the plasma flow velocity $\bs{v}$ and its vorticity, such that measuring canonical helicity generally requires multiple local, 3D velocity measurements.
Another key ingredient is the cross helicity density $\bs{v}\cd\bs{B}$,
\cite{you_2012, you_2014, von_der_linden_2018},
where $\bs{B}$ is the magnetic field, also motivating local, 3D velocity measurements.

Various diagnostics have been used to measure local plasma flows in RFPs,
including ion Doppler spectroscopy (IDS) probes
\cite{fiksel_1998, den_hartog_1999, fontana_1999, fontana_2000, kuritsyn_2008, kuritsyn_2009, miller_2011},
localized active IDS measurements with charge-exchange recombinations spectroscopy (CHERS)
\cite{craig_2007, ennis_2010, craig_2019, boguski_2021},
and 2D Mach probes
\cite{antoni_1996b, spolaore_2002, kuritsyn_2008, kuritsyn_2009, de_masi_2010, miller_2011}.
Compared to spectroscopic methods, Mach probes
\cite{chung_2012}
offer relatively simple operational methodology and also lend themselves to convenient combinations with other probe-based diagnostics, while the translation of raw Mach probe signals into velocity measurements can be challenging depending on the experimental scenario.

In this paper, a novel matrix method for analyzing raw data from a general 3D unmagnetized Mach probe array is developed and used to provide velocity values for two 3D Mach probe arrays
\cite{sellner_2024, rouda_2025}
used for plasma flow measurement testing in the edge of RFP plasmas in the Madison Symmetric Torus (MST)
\cite{dexter_1991}.
The matrix method facilitates analysis of 3D Mach probe arrays in which the probe tips are not necessarily composed of pairs of mutually opposite tips.
The two 3D Mach probes tested were composed of regular polyhedral arrays of six and four tips, respectively, with the six-tip array composed of three orthogonal pairs of mutually opposite tips at the vertices of a regular octahedron
\cite{sellner_2024}
and the four-tip array composed of non-opposite tips at the vertices of a regular tetrahedron
\cite{rouda_2025}.
Analysis of the tetrahedral array data is specifically facilitated by the general matrix method.
We also calculate measurement uncertainties for the Mach probes based on machining precision and ion saturation current measurement uncertainties, estimating it to be of order several percent, smaller than the systematic uncertainty in Mach number calibration constant and experimental uncertainties related to plasma and probe conditioning.

We qualitatively compare our initial flow measurements to expections based on previous MST results
\cite{fontana_1999, miller_2011}.
Our initial octahedron probe results show equilibrium flow speeds roughly as expected, and the flow direction was similar to that in previous 2D Mach probe results
\cite{miller_2011},
while the overall flow direction was roughly the opposite of the expected E-cross-B direction
\cite{fontana_1999}.
Some of the initial tetrahedron results show roughly the expected E-cross-B toroidal and poloidal flow directions but involve larger than expected speeds, especially in the radial direction.
We consider possible causes of the lack of clear agreement, focusing on possible probe conditioning and fast electron
\cite{stoneking_1994}
issues.

This paper is organized as follows.
The 3D Mach probe analysis is presented in Sec.\ \ref{sec:analysis}, with Mach probe background in Sec.~\ref{subsec:mach_probes}, the matrix method developed in Sec.~\ref{subsec:matrix_method}, application to a six-tipped regular octahedral probe in Sec.~\ref{subsec:octa_probe} and four-tipped regular tetrahedral probe in Sec.~\ref{subsec:tetra_probe}, and measurement uncertainty estimates in Sec.~\ref{subsec:unc_estimates}.
The experiments performed for this paper are described in
Sec.\ \ref{sec:experiments}.
The results are presented in Sec.\ \ref{sec:results}, specifically for the octahedral probe in Sec.~\ref{subsec:octa_results} and the tetrahedral probe in Sec.~\ref{subsec:tetra_results}.
These topics are summarized and discussed in Sec.\ \ref{sec:discussion}.

\section{Analysis}
\label{sec:analysis}

\subsection{Mach probe background}
\label{subsec:mach_probes}

Mach probes
\cite{chung_2012}
are a well-established diagnostic in experimental plasma physics to measure plasma flow velocity by collecting ion saturation currents from multiple negatively biased Langmuir probe tips.
The basic case is a one-dimensional (1D) Mach probe composed of a pair of such tips oriented oppositely, 180$^\circ$ apart from each other, with an insulating probe housing between them.
For an unmagnetized probe, which has tip size significantly smaller than the ion cyclotron (Larmor) radius, the projection $v$ of the plasma flow velocity $\bs{v}$ along the tip pair axis can then be related to the upstream and downstream ion saturation currents $I_\T{u}$ and $I_\T{d}$ by
\begin{equation}
  \label{eq:up-down_mach}
  I_\T{u}/I_\T{d} = \exp\lp Kv/v_\T{s}\rp = \exp\lp KM\rp,
\end{equation}
where $K$ is a calibration constant depending on the plasma model, $v_\T{s}$ is a suitably defined ion sound or thermal speed,
and $M \equiv v/v_\T{s}$ is the Mach number of the projected velocity
\cite{chung_2006}.

A Mach probe model derived from kinetic plasma simulations by Hutchinson
\cite{hutchinson_2002c}
applies to an unmagnetized plasma with zero Debye length $\lambda_\T{D}$, and a key result is a calculated calibration constant $K = 1.34$, with a corresponding definition for normalizing ion sound speed for the Mach number of $v_\T{s}\equiv\sqrt{k_\T{B}Z\Te/m_\T{i}}$, where $k_\T{B}$ is Boltzmann's constant, $Z$ is the ion charge, $\Te$ is the electron temperature, and $m_\T{i}$ is the ion mass.

The 1D Mach probe relationship in Eq.~(\ref{eq:up-down_mach}) has often been generalized to 2D arrays with two orthogonal pairs of Mach tips
\cite{antoni_1996b, kuritsyn_2008, kuritsyn_2009, miller_2011}
or to 2D arrays with multiple tips in a ring pattern called `Gundestrup' probes
\cite{maclatchy_1992, spolaore_2002, de_masi_2010, spolaore_2011, scarin_2011, roche_2014},
typically allowing simultaneous measurement of poloidal and toroidal flow velocity components.
An important aspect of such measurements
\cite{peterson_1994, antoni_1996b, roche_2014} is that the ion saturation current $I$ collected by a planar Mach tip electrode face depends on the direction cosine $\cos\delta$ of the angle $\delta$ between the flow velocity and electrode normal directions as in
\begin{align}
  \begin{split}
    \label{eq:hutch_2002c_13}
    I &= I_0
    \exp\lc\lp M/2\rp\right.
    \\
    &\x \left.\ls K_\T{u}\lp 1 - \cos\delta\rp
    - K_\T{d}\lp 1 + \cos\delta\rp\rs\rc,
  \end{split}
\end{align}
where $I_0$ is the level without flow, and the calibration constants $K_\T{u}$ and $K_\T{d}$ satisfy $K = K_\T{u} + K_\T{d}$
\cite{hutchinson_2002c, roche_2014}.

\subsection{Matrix method}
\label{subsec:matrix_method}

A novel aspect of the present work is to generalize the considerations for 2D Mach probes to 3D, including Mach tips that are not necessarily directly opposite to each other.
This not only enables a six-tip Mach probe with three orthogonal pairs of directly opposite tips but also a four-tip tetrahedral probe with no two tips directly opposite to each other.
In principle, the four-tip approach allows a more compact 3D probe than the six-tip, which could be an important advantage in experimental contexts where Mach probe size is a key issue, as it tends to be in MST, with its close-fitting conducting shell and limited port access.

With two arbitrarily oriented tips $i$ and $j$, Eq.~(\ref{eq:hutch_2002c_13}) becomes
\begin{align}
  \label{eq:two_arb_tips_ln}
  I_i/I_j &= \exp\ls -\lp KM/2\rp\lp\cos\delta_i - \cos\delta_j\rp\rs.
\end{align}
For a 1D upstream-downstream Mach probe with a pair of oppositely aligned tips respectively labeled $i=\T{u}$ and $j=\T{d}$, with
$\cos\delta_\T{u} = -1$ and $\cos\delta_\T{d} = 1$,
this recovers Eq.~(\ref{eq:up-down_mach}).

In three dimensions (3D), the direction cosine for a tip labeled `$i$' with normal unit vector
\begin{align*}
  \whbs{n}_i = \bxi\whbs{x} + \byi\whbs{y} + \bzi\whbs{z}
\end{align*}
with the flow velocity unit vector
\begin{align*}
  \whbs{v} = \bvx\whbs{x} + \bvy\whbs{y} + \bvz\whbs{z}
\end{align*}
is
\begin{align}
  \begin{split}
    \label{eq:dir_cos}
    \cos\delta_i &= \whbs{n}_i \cd \whbs{v}
    = \bxi\bvx + \byi\bvy + \bzi\bvz,
  \end{split}
\end{align}
where
the overbar symbol ``$\bar{\phantom{x}}$'' indicates a component of a unit vector,
$\whbs{x}$, $\whbs{y}$, and $\whbs{z}$ are the coordinate unit vectors in a Cartesian coordinate system,
and the components of $\whbs{n}_i$ are
\begin{align}
  \begin{split}
    \label{eq:xyz_theta_phi}
    \bxi &= \sin\theta_i\cos\phi_i,
    \\
    \byi &= \sin\theta_i\sin\phi_i,
    \\
    \bzi &= \cos\theta_i,
  \end{split}
\end{align}
where $\theta_i$ and $\phi_i$ are the polar and azimuthal angles of tip `$i$' in a spherical coordinate system.

A matrix problem for a 3D probe can then be built up by populating a $3\x3$ matrix equation with elemental relations for each tip combination in the probe.
Each is a linear combination of differences of direction cosines for one or more not necessarily directly opposing tip pairs related to the tip current ratios.
Combining Eqs.~(\ref{eq:two_arb_tips_ln}) and (\ref{eq:dir_cos}), each difference is given by
\begin{align*}
    &M\lp\whbs{n}_i - \whbs{n}_j\rp\cd\whbs{v} 
    \\
    &= M\ls\lp\bxi - \bxj\rp\bvx
      + \lp\byi - \byj\rp\bvy
      + \lp\bzi - \bzj\rp\bvz\rs
    \\
    &= -\lp 2/K\rp\ln\lp I_i/I_j\rp,
\end{align*}
with one matrix row for each tip combination $k$, $l$, or $m$,
\begin{align*}
  k: (i, j) &= \lp\T{a}, \T{b}\rp, \lp\T{c}, \T{d}\rp, \ldots ;
  \\
  l: (i, j) &= \lp\T{e}, \T{f}\rp, \lp\T{g}, \T{h}\rp, \ldots ;
  \\
  m: (i, j) &= \lp\T{p}, \T{q}\rp, \lp\T{r}, \T{s}\rp, \ldots.
\end{align*}
Then the general matrix problem is, with $\xk \equiv \sum_k\lp\bxi - \bxj\rp$, \etc.,
\begin{align}
  \begin{split}
      \label{eq:general_matrix_equation}
      &M
      \begin{pmatrix}
        \xk & \yk & \zk \\
        \xl & \yl & \zl \\
        \xm & \ym & \zm
      \end{pmatrix}
      \begin{pmatrix}
        \bvx \\
        \bvy \\
        \bvz
      \end{pmatrix}
      \\
      &= -\fr{2}{K}
      \begin{Bmatrix}
        \ln\ls\Pi_k\lp I_i/I_j\rp\rs \\
        \ln\ls\Pi_l\lp I_i/I_j\rp\rs \\
        \ln\ls\Pi_m\lp I_i/I_j\rp\rs
    \end{Bmatrix}
  \end{split}
\end{align}
or
\begin{align}
  \label{eq:short_form}
  \bs{A}\fr{\bs{v}}{v_\T{s}} &= -\fr{2}{K}\bs{J},
\end{align}
where
\begin{align*}
  \bs{A} = 
  \begin{pmatrix}
    \xk & \yk & \zk \\
    \xl & \yl & \zl \\
    \xm & \ym & \zm
  \end{pmatrix},
\end{align*}
\begin{align*}
  \fr{\bs{v}}{v_\T{s}} = 
  M
  \begin{pmatrix}
    \bvx \\
    \bvy \\
    \bvz
  \end{pmatrix},
\end{align*}
and
\begin{align*}
  \bs{J} = 
  \begin{Bmatrix}
    \ln\ls\Pi_k\lp I_i/I_j\rp\rs \\
    \ln\ls\Pi_l\lp I_i/I_j\rp\rs \\
    \ln\ls\Pi_m\lp I_i/I_j\rp\rs
  \end{Bmatrix}.
\end{align*}
If $\bs{A}$ is invertible, then the solution is
\begin{align}
  \label{eq:general_solution}
  \fr{\bs{v}}{v_\T{s}}
  &=
  -\fr{2}{K}\bs{A}^{-1}\bs{J}.
\end{align}
If $\bs{A}$ has orthogonal rows with all equal norms
$w = \sqrt{x_t^2 + y_t^2 + z_t^2}$ for $t = k, l, m$,
then the inverse is
\begin{align*}
  \bs{A}^{-1}
  &= 
  \fr{1}{w^2}
  \begin{pmatrix}
    \xk & \xl & \xm \\
    \yk & \yl & \ym \\
    \zk & \zl & \zm
  \end{pmatrix}
  = 
  \fr{1}{w}
  \begin{pmatrix}
    \bxk & \bxl & \bxm \\
    \byk & \byl & \bym \\
    \bzk & \bzl & \bzm
  \end{pmatrix},
\end{align*}
where $\bxk$, \etc. are the unit vector components of the linear combinations of tip directions.

\subsection{Octahedral probe}
\label{subsec:octa_probe}

\begin{figure}
\centering
  \includegraphics[width=\columnwidth]{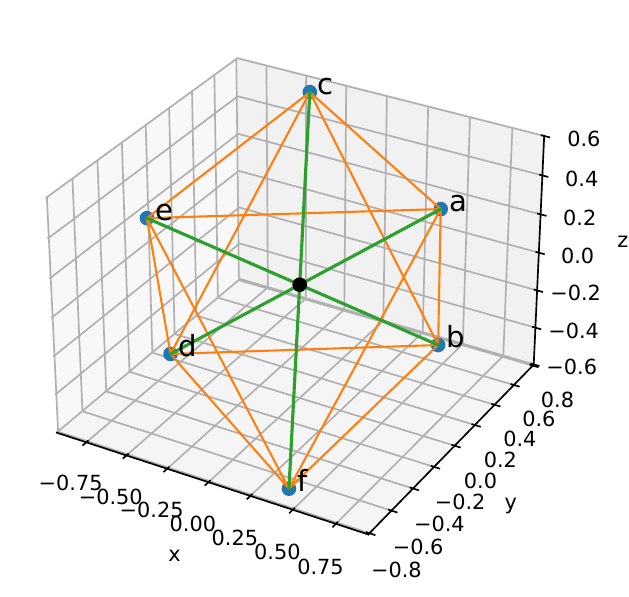}
  \caption{\label{fig:octa_example}Octahedral probe example.}
\end{figure}

A six-tipped `octahedral' probe has Mach tips `a' through `f' at the six vertices of a regular octahedron, as shown in the example in FIG.~\ref{fig:octa_example} for the set of probes angles
\begin{align}
  \begin{split}
    \label{eq:octa_angles}
    \thea, \phia &= \atsrt\degs, 0\degs
                  = \arctan\lp\sqrt{2}\rp, 0; \\
    \theb, \phib &= \tpmatsrt\degs, 60\degs
                  = \pi - \arctan\lp\sqrt{2}\rp, \pi/3; \\
    \thec, \phic &= \atsrt\degs, 120\degs; \\
    \thed, \phid &= \tpmatsrt\degs, 180\degs; \\
    \thee, \phie &= \atsrt\degs, 240\degs; \\
    \thef, \phif &= \tpmatsrt\degs, 300\degs.
  \end{split}
\end{align}

\noindent

A convenient set of tip combinations in this case is composed of the three pairs of opposing tips with mutually orthogonal axes.
With the definitions $\xad \equiv \bxa - \bxd$, \etc., the matrix problem Eq.~(\ref{eq:general_matrix_equation}) becomes
\begin{align*}
  M
  \begin{pmatrix}
    \xad & \yad & \zad \\
    \xbe & \ybe & \zbe \\
    \xcf & \ycf & \zcf
  \end{pmatrix}
  \begin{pmatrix}
    \bvx \\
    \bvy \\
    \bvz
  \end{pmatrix}
  &= -\fr{2}{K}
  \begin{bmatrix}
    \ln\lp\Ia/\Id\rp \\
    \ln\lp\Ib/\Ie\rp \\
    \ln\lp\Ic/\If\rp
  \end{bmatrix},
\end{align*}
with $\bs{J}$ in Eq.~(\ref{eq:short_form}) given by
\begin{align*}
  \bs{J}_\T{oct} &= 
  \begin{bmatrix}
    \ln\lp\Ia/\Id\rp \\
    \ln\lp\Ib/\Ie\rp \\
    \ln\lp\Ic/\If\rp
  \end{bmatrix}
\end{align*}
and $\bs{A}$ in Eq.~(\ref{eq:short_form}) given by
\begin{align*}
  \bs{A}_\T{oct} &= 
  \begin{pmatrix}
    \xad & \yad & \zad \\
    \xbe & \ybe & \zbe \\
    \xcf & \ycf & \zcf
  \end{pmatrix}
  =
  2
  \begin{pmatrix}
    \bxa & \bya & \bza \\
    \bxb & \byb & \bzb \\
    \bxc & \byc & \bzc
  \end{pmatrix},
\end{align*}
whose elements can be expressed numerically via Eq.~(\ref{eq:xyz_theta_phi}) using the angles exemplified in Eq.~(\ref{eq:octa_angles}).
This matrix has orthogonal rows and columns with all equal norms
$\sqrt{\xij^2 + \yij^2 + \zij^2} = 2$ for $ij = \T{ad}, \T{be}, \T{cf}$.
The matrix inverse is
\begin{align*}
  \bs{A}_\T{oct}^{-1}
  &=
  \fr{1}{4}
  \begin{pmatrix}
    \xad & \xbe & \xcf \\
    \yad & \ybe & \ycf \\
    \zad & \zbe & \zcf
  \end{pmatrix}
  =
  \fr{1}{2}
  \begin{pmatrix}
    \bxa & \bxb & \bxc \\
    \bya & \byb & \byc \\
    \bza & \bzb & \bzc
  \end{pmatrix},
\end{align*}
such that the solution as in Eq.~(\ref{eq:general_solution}) is
\begin{align}
  \begin{split}
    \label{eq:octa_solution}
    \fr{\bs{v}}{v_\T{s}}
    &= -\fr{1}{2K}
    \begin{pmatrix}
      \xad & \xbe & \xcf \\
      \yad & \ybe & \ycf \\
      \zad & \zbe & \zcf
    \end{pmatrix}
    \bs{J}_\T{oct}
    \\
    &= -\fr{1}{K}
    \begin{pmatrix}
      \bxa & \bxb & \bxc \\
      \bya & \byb & \byc \\
      \bza & \bzb & \bzc
    \end{pmatrix}
    \begin{bmatrix}
      \ln\lp\Ia/\Id\rp \\
      \ln\lp\Ib/\Ie\rp \\
      \ln\lp\Ic/\If\rp
    \end{bmatrix},
  \end{split}
\end{align}
or, for the example in Eq.~(\ref{eq:octa_angles}) and FIG.~\ref{fig:octa_example},
\begin{align*}
  \fr{\bs{v}}{v_\T{s}}
  &= -\fr{1}{K}
  \begin{pmatrix}
    0.81650 &  0.40825 & -0.40825 \\
    0.0     &  0.70711 &  0.70711 \\
    0.57735 & -0.57735 &  0.57735
  \end{pmatrix}
  \bs{J}_\T{oct}.
\end{align*}
As suggested by comparison of the last expression in Eq.~(\ref{eq:octa_solution}) with Eq.~(\ref{eq:up-down_mach}), it is straightforward to show that this octahedral probe solution is a composition of the results from the three orthogonal 1D upstream-downstream Mach tip pairs that make up the probe.

\subsection{Tetrahedral probe}
\label{subsec:tetra_probe}

\begin{figure}
\centering
  \includegraphics[width=\columnwidth]{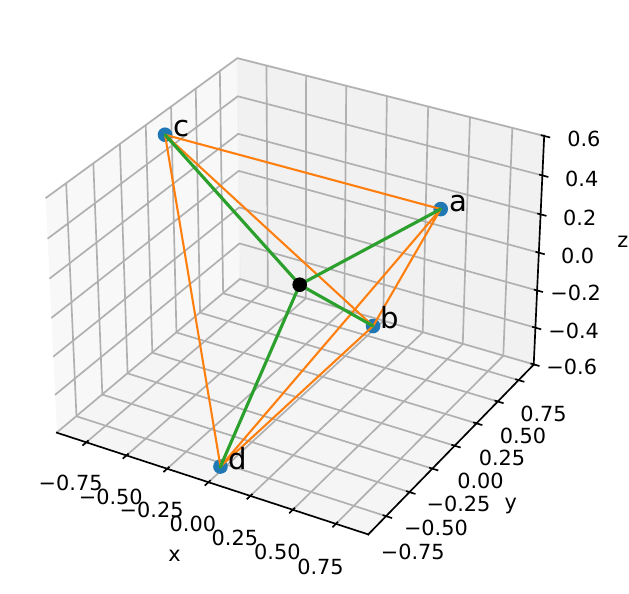}
  \caption{\label{fig:tetra_example}Tetrahedral probe example.}
\end{figure}

A four-tipped tetrahedral probe has Mach tips `a' through `d' at the vertices of a regular tetrahedron, as shown in the example in FIG.~\ref{fig:tetra_example} for the set of probes angles
\begin{align}
  \begin{split}
    \label{eq:tetra_angles}
    \thea, \phia &= \atsrt\degs, 0\degs
                  = \arctan\lp\sqrt{2}\rp, 0; \\
    \theb, \phib &= \tpmatsrt\degs, 90\degs
                  = \pi - \arctan\lp\sqrt{2}\rp, \pi/2; \\
    \thec, \phic &= \atsrt\degs, 180\degs; \\
    \thed, \phid &= \tpmatsrt\degs, 270\degs. \\
  \end{split}
\end{align}

\noindent

Unlike the case for an octahedral probe, for a tetrahedral probe, the general 3D matrix solution method in Sec.~\ref{subsec:matrix_method} is not equivalent to considering pairs of directly opposing \textit{tips}, since the tetrahedral arrangement involves no such tips.
In this case, a convenient set of tip combinations is composed of the three successive sets of opposing \textit{pairs} of tips with mutually orthogonal average directions.
With the definitions $\xabcd \equiv \bxa + \bxb - \bxc - \bxd$, \etc., entailing a different notation convention than that in Sec.~\ref{subsec:octa_probe}, the matrix problem Eq.~(\ref{eq:general_matrix_equation}) becomes
\begin{align*}
  &M
  \begin{pmatrix}
    \xabcd & \yabcd & \zabcd \\
    \xbcda & \ybcda & \zbcda \\
    \xacbd & \yacbd & \zacbd
  \end{pmatrix}
  \begin{pmatrix}
    \bvx \\
    \bvy \\
    \bvz
  \end{pmatrix}
  \\
  &= -\fr{2}{K}
  \begin{Bmatrix}
    \ln\ls\lp\Ia\Ib\rp/\lp\Ic\Id\rp\rs \\
    \ln\ls\lp\Ib\Ic\rp/\lp\Id\Ia\rp\rs \\
    \ln\ls\lp\Ia\Ic\rp/\lp\Ib\Id\rp\rs
  \end{Bmatrix},
\end{align*}
with $\bs{J}$ in Eq.~(\ref{eq:short_form}) given by
\begin{align*}
  \bs{J}_\T{tet} &= 
  \begin{Bmatrix}
    \ln\ls\lp\Ia\Ib\rp/\lp\Ic\Id\rp\rs \\
    \ln\ls\lp\Ib\Ic\rp/\lp\Id\Ia\rp\rs \\
    \ln\ls\lp\Ia\Ic\rp/\lp\Ib\Id\rp\rs
  \end{Bmatrix}
\end{align*}
and $\bs{A}$ in Eq.~(\ref{eq:short_form}) given by
\begin{align*}
  \bs{A}_\T{tet} &= 
  \begin{pmatrix}
    \xabcd & \yabcd & \zabcd \\
    \xbcda & \ybcda & \zbcda \\
    \xacbd & \yacbd & \zacbd
  \end{pmatrix},
\end{align*}
whose elements can be expressed numerically via Eq.~(\ref{eq:xyz_theta_phi}) using the angles exemplified in Eq.~(\ref{eq:tetra_angles}).
This matrix has orthogonal rows and columns with all equal norms $\sqrt{x_{ijkl}^2 + y_{ijkl}^2 + z_{ijkl}^2} = 4/\sqrt{3}$ for $ijkl = \T{abcd}, \T{bcda}, \T{acbd}$.
The matrix inverse $\bs{A}^{-1}$ is
\begin{align*}
  \bs{A}_\T{tet}^{-1}
  &=
  \fr{3}{16}
  \begin{pmatrix}
    \xabcd & \xbcda & \xacbd \\
    \yabcd & \ybcda & \yacbd \\
    \zabcd & \zbcda & \zacbd
  \end{pmatrix},
\end{align*}
such that the solution as in Eq.~(\ref{eq:general_solution}) is
\begin{align}
  \begin{split}
    \label{eq:tetra_solution}
    \fr{\bs{v}}{v_\T{s}}
    &= -\fr{3}{8K}
    \begin{pmatrix}
      \xabcd & \xbcda & \xacbd \\
      \yabcd & \ybcda & \yacbd \\
      \zabcd & \zbcda & \zacbd
    \end{pmatrix}
    \bs{J}_\T{tet},
  \end{split}
\end{align}
or, for the example in Eq.~(\ref{eq:tetra_angles}) and FIG.~\ref{fig:tetra_example},
\begin{align*}
  \fr{\bs{v}}{v_\T{s}}
  &= -\fr{1}{K}
  \begin{pmatrix}
    0.61237 & -0.61237 & 0.0 \\
    0.61237 &  0.61237 & 0.0 \\
    0.0     &  0.0     & 0.86603
  \end{pmatrix}
  \bs{J}_\T{tet}.
\end{align*}

\subsection{Measurement uncertainty estimates}
\label{subsec:unc_estimates}

Given the solutions in Eq.~(\ref{eq:octa_solution}) and Eq.~(\ref{eq:tetra_solution}) for octahedral and tetrahedral Mach probes, the expected measurement uncertainties can be estimated by standard error propagation techniques
\cite{bevington_2003}
assuming normally distributed errors in probe machining and ion saturation current measurements.

Velocity measurement uncertainty $\sigma_{v,xyz}$ for each vector component due to probe machining uncertainty can be estimated via derivatives of the $\bs{A}$ matrix with respect to the spatial coordinates $w_i \equiv \xa, \ya, \za, \xb, \ldots$ combined with an assumed overall spatial uncertainty $\sigma_{xyz}$ in the tip positions, as in
\begin{align}
  \begin{split}
    \label{eq:unc_xyz_deriv}
    \sigma_{v,xyz}^2
    &=
    \sigma^2_{xyz}\sum_{w_i} \lp \fr{\pd\bs{v}}{\pd w_i}\rp^2 
    \\
    &=
    \sigma^2_{xyz}\sum_{w_i} \ls \bs{A}^{-1}\lp \fr{\pd\bs{A}}{\pd w_i}\rp\bs{v} \rs^2,
  \end{split}
\end{align}
which is in the form of a column array with all identical entries.
Equation~(\ref{eq:unc_xyz_deriv}) reflects a model of the probe machining errors modifying the otherwise ideal measured saturation current signals ($\bs{J}\propto\bs{A}\bs{v}$) by which the unchanged matrix inverse is multiplied to obtain the velocity measurement.

Velocity measurement uncertainty $\sigma_{v,I}$ due to saturation current measurement uncertainty, \eg.\ that ascribed to uncertainty of resistor values in voltage dividers, can be estimated via derivatives of the $\bs{A}$ matrix with respect to the saturation currents $I_i$ for $i \equiv \T{a}, \T{b}, \T{c}, \ldots$ combined with an assumed saturation current uncertainty $\sigma_I$, as in the column-array form
\begin{align}
  \begin{split}
    \label{eq:unc_isat_deriv}
    \sigma_{v,I}^2
    &=
    \sigma^2_{I}\sum_{I_i} \lp \fr{\pd\bs{v}}{\pd I_i}\rp^2 
    \\
    &=
    \sigma^2_{I}\sum_{I_i} \lp \fr{-2v_\T{s}\bs{A}^{-1}}{K}\fr{\pd\bs{J}}{\pd I_i}\rp^2
    \\
    &=
    4\lp\fr{v_\T{s}}{K}\rp^2\sigma^2_{I}\sum_{I_i} \lp \bs{A}^{-1}\fr{\pd\bs{J}}{\pd I_i}\rp^2.
  \end{split}
\end{align}

Using Eqs.~(\ref{eq:unc_xyz_deriv}) and (\ref{eq:unc_isat_deriv}), we calculate the sums of the squares of these uncertainties to estimate the relative measurement uncertainties for the octahedral probe case, given by
\begin{align}
  \begin{split}
    \label{eq:octa_meas_unc}
    \lp\fr{\sigma_{v,\T{oct}}}{v}\rp^2
    &=
    \fr{1}{2}\lp\fr{\sigma_{xyz}}{r}\rp^2 + \fr{2}{\lp KM\rp^2}\lp\fr{\sigma_{I}}{I}\rp^2,
  \end{split}
\end{align}
and for the tetrahedral probe case, given by
\begin{align}
  \begin{split}
    \label{eq:tetra_meas_unc}
    \lp\fr{\sigma_{v,\T{tet}}}{v}\rp^2
    &=
    \fr{3}{4}\lp\fr{\sigma_{xyz}}{r}\rp^2 + \fr{3}{\lp KM\rp^2}\lp\fr{\sigma_{I}}{I}\rp^2,
  \end{split}
\end{align}
where $r$ is the probe tip radius, and $I$ is a typical ion saturation current value, such that $\sigma_\T{xyz}/r$ and $\sigma_I/I$ are relative uncertainties for probe machining and ion saturation current measurement, respectively.
We note that the estimates in Eqs.~(\ref{eq:octa_meas_unc}) and (\ref{eq:tetra_meas_unc}) imply that the square of the ratio of measurement uncertainties for the two probe types is $\lp\sigma_{v,\T{tet}}/\sigma_{v,\T{oct}}\rp^2 = 3/2$, which is the inverse of the ratio of numbers of Mach tips composing the probes.
This makes intuitive sense in that more tips collect more information.
From Eqs.~(\ref{eq:octa_meas_unc}) and (\ref{eq:tetra_meas_unc}), given expected values of a few percent each for $\sigma_\T{xyz}/r$ and $\sigma_I/I$, we expect relative velocity measurement uncertainties of $\sigma_v/v$ around several percent, which is likely smaller than either the systematic uncertainty in choosing a value of $K$ for an unmagnetized plasma
\cite{chung_2012, hutchinson_2003}
or the experimental uncertainties of plasma and probe conditioning.

\section{Experiments}
\label{sec:experiments}

MST
\cite{dexter_1991}
is a toroidal device with major radius $R=1.5$~m, minor radius $a=52$~cm, and a 5~cm thick Al shell as a flux conserver.
Our Mach probe experiments
\cite{sellner_2024, rouda_2025}
involved probe insertion depths of up to 10~cm into the plasma edge, via a 3.8~cm diameter port on the top side of the flux conserver, for RFP shots with $\Ip$ values of up to about 200~kA and pulse durations of about 60~ms, during which the plasma is expected to exhibit spontaneous equilibrium flows and 3D flow fluctuations as described in Sec.~\ref{sec:intro}.

\begin{figure}
\centering
  \includegraphics[width=\columnwidth]{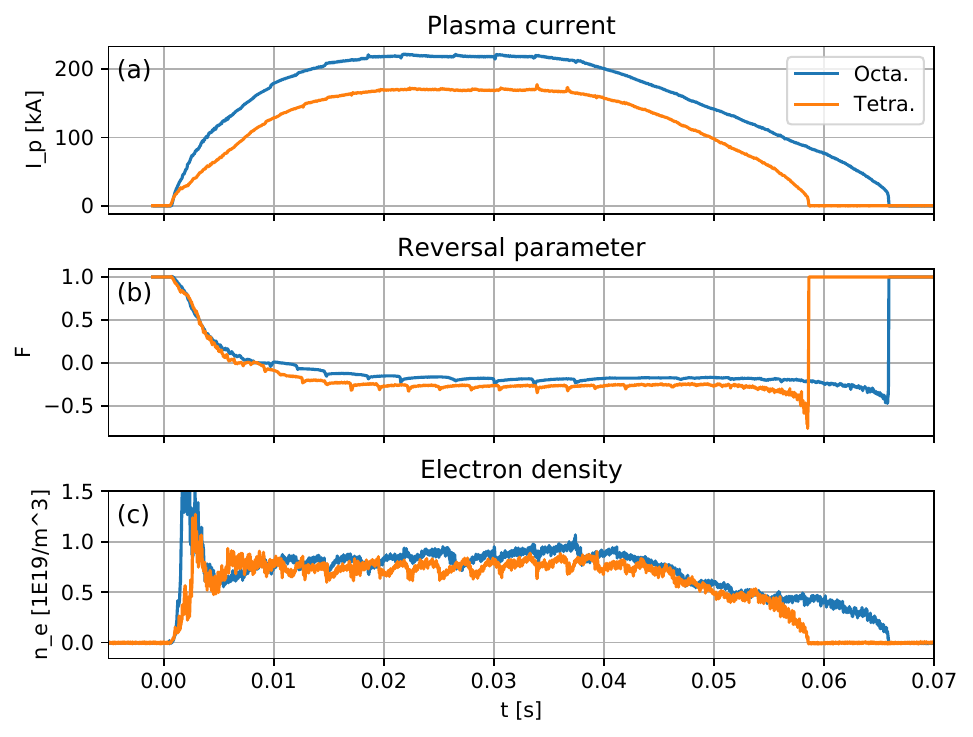}
  \caption{\label{fig:mst_shots}RFP operational signals vs.\ time for the example shot with the octahedral probe in blue (see FIG.~\ref{fig:mst_octa}), tetrahedral probe in orange (see FIG.~\ref{fig:mst_tetra}). (a) Plasma current $\Ip$.  (b) Reversal parameter $F$.  (c) Electron density $\nel$.}
\end{figure}

Key operational signals for the RFP plasmas used for both the octahedral and tetrahedral Mach probe experiments are shown in FIG.~\ref{fig:mst_shots}.
These are (a) $\Ip$, (b) the RFP equilibrium reversal parameter $F\equiv B_\T{t,wall}/\la\Bt\ra$, which is the ratio of $\Bt$ at the wall and the cross-section average $\Bt$, with typical value -0.2, and (c) the electron density $\nel$ with typical value $0.8\x10^{19}/\T{m}^3$.
Note the sawtooth oscillations at a few hundred Hz most visible on the $F$ and $\nel$ signals, indicating magnetic relaxation events that could be a focus of canonical helicity studies
\cite{you_2012}.

Around fifty to a hundred shots were taken per run day.
To promote stable plasma conditions, the probe was first gradually inserted with a few shots per step until it reached its target depth.
Otherwise, if inserted too quickly, the plasma discharge quality could suffer, preventing useful experiments.
The negative bias applied to the probe tips was started low and then gradually increased until the measured currents stopped changing with larger bias, indicating ion saturation.
The tip current signals were measured with resistive voltage dividers and digitized, with sampling rates of 1~MHz for the octahedral probe and 200~kHz for the tetrahedral probe.
On some run days, the probe was rotated on its axis throughout the day, typically every dozen shots or so, collecting information about possible tip dependencies in the flow measurement process.

\noindent

Ion saturation current signals sometimes exhibited large excursions indicating plasma arcing that gradually diminished in frequency and amplitude as the probe condition improved, especially during the early stages of a run.
Fast electrons
\cite{stoneking_1994}
traveling in the negative poloidal direction (upward on the outboard side, \ie.\ oppositely to the poloidal current density) are also known to be a possible cause of noise in the saturation current signal.

Experiments with the octahedral probe
\cite{sellner_2024}
were performed over a few run days, with most of the shots taken on a single run day, after which the Mach tip connections internal to the probe were found to have been significantly damaged by heat from the plasma, and the probe was withdrawn from service.
The tetrahedral probe
\cite{rouda_2025}
experiments described in this work used a probe with four stalks designed to measured local canonical vorticity, such that each probe stalk supported a tetrahedral Mach probe as described in Sec.~\ref{subsec:tetra_probe}.
One of the four tetrahedral Mach probes in the vorticity probe survived for several days of experiments, which produced the results discussed here.
By the end of this run campaign, the vorticity probe had become heavily damaged by the plasma and was withdrawn from service.

\section{Results}
\label{sec:results}

The key numerical inputs and coordinate conventions used in deriving the measurement results of these experiments are described, and example 3D velocity measurements for each probe are presented.
We use the Mach probe calibration constant $K=1.34$ for convenience, as both the unmagnetized and zero-$\lambda_\T{D}$ assumptions involved in its derivation
\cite{hutchinson_2002c}
are approximately satisfied in our MST plasmas, and it has been used in a previous Mach probe study on MST
\cite{miller_2011}
to whose results we will compare ours in Sec.~\ref{sec:discussion}.
To estimate the sound speeds for the Mach probe analysis at the relevant probe insertion depths and thereby express the flow velocity measurements with units of km/s, we use previously measured $\Ip$ and $\nel$ scalings of core $\Te$ values
\cite{stoneking_1998, jacobson_2017b, kubala_2023}
and edge $\Te$ profiles
\cite{kubala_2023}
indexed to the individual probe shot $\Ip$ and $\nel$ values and insertion depths.

Depending on the probe rotation value for each shot, we translate between the matrix coordinate system in Sec.~\ref{sec:analysis} and the standard MST right-handed $(r, \theta, \phi)$ coordinate system, where $\whbs{r}$ is in the minor radial direction, $\whbs{\theta}$ is in the poloidal direction pointing outboard on the top side of the torus, and $\whbs{\phi}$ is in the toroidal direction pointing counterclockwise (CCW) when viewed from the top.
In MST, $\Ip$ is typically directed clockwise (CW), opposite to the CCW core $\Bt$.

\begin{figure}
\centering
  \includegraphics[width=\columnwidth]{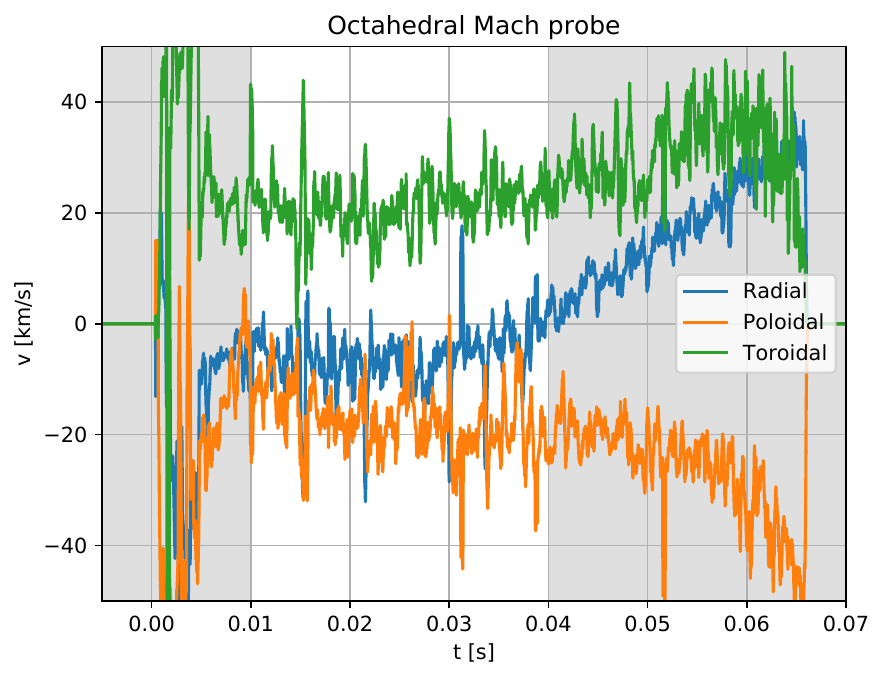}
  \caption{\label{fig:mst_octa}Initial measurements of 3D plasma velocity components vs.\ time from the octahedral Mach probe inserted 4.5~cm deep into the plasma for the blue shot in FIG.~\ref{fig:mst_shots}.  Radial component in blue; positive means outward.  Poloidal component in orange; positive means toward outboard on top.  Toroidal component in green; positive means counterclockwise (CCW) from above, with $\Bt$, against $\Ip$.  White time window indicates the plasma flattop; shaded time windows indicate plasma startup and rampdown.  Data is boxcar-averaged over a running time window of 250 $\mu$s.}
\end{figure}\noindent

\subsection{Octahedral probe results}
\label{subsec:octa_results}

The octahedral probe provided a series of about ten shots with stable ion saturation current signal behavior and without evidence of significant arcing.
The probe insertion depth was 4.5~cm, and the bias voltage applied to the tips was about 5.5 times larger than the estimated local $\Te$.
An example of an initial measurement of 3D plasma velocity from one of these shots is shown in FIG.~\ref{fig:mst_octa}.
The toroidal component is shown in green, the poloidal in orange, and the radial in blue.
During the RFP flattop period indicated with the white background in the plot, the toroidal and poloidal components are each around 20~km/s in magnitude, with the toroidal in the positive $\whbs{\phi}$ direction and the poloidal in the negative $\whbs{\theta}$ direction.
The radial component is smaller, around 5~km/s, and in the negative $\whbs{r}$ direction.
The measurements for the plasma startup and rampdown phases, indicated with the gray backgrounds in the plot, exhibit faster changes and larger magnitudes than do those for the flattop.

\subsection{Tetrahedral probe results}
\label{subsec:tetra_results}

\begin{figure}
\centering
  \includegraphics[width=\columnwidth]{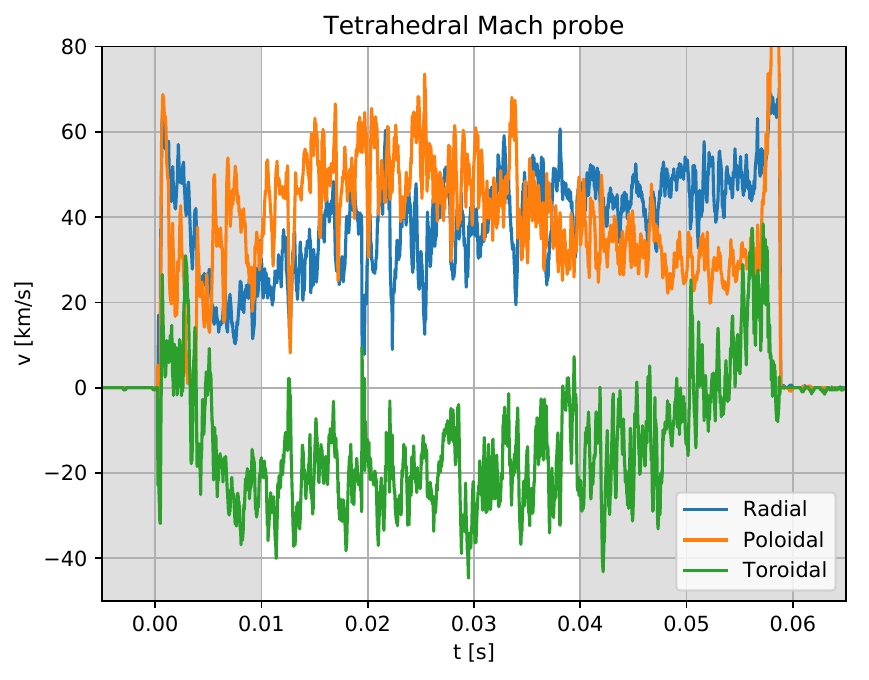}
  \caption{\label{fig:mst_tetra}Initial measurements of 3D plasma velocity components vs.\ time from the tetrahedral Mach probe inserted 7~cm deep into the plasma for the orange shot in FIG.~\ref{fig:mst_shots}.  Radial component in blue; positive means outward.  Poloidal component in orange; positive means toward outboard on top.  Toroidal component in green; positive means CCW from above, with $\Bt$, against $\Ip$.  White time window indicates the plasma flattop; shaded time windows indicate plasma startup and rampdown.  Data is boxcar-averaged over a running time window of 250 $\mu$s.}
\end{figure}

The tetrahedral probe provided one series of about twenty shots with more stable ion saturation current signal behavior compared to others and without evidence of significant arcing.
The probe insertion depth was 7~cm, and the bias voltage applied to the tips was about 3 times larger than the estimated local $\Te$.
An example of an initial measurement of 3D plasma velocity from one of these shots is shown in FIG.~\ref{fig:mst_tetra}.
During the RFP flattop period,
the magnitude of the toroidal component is around 20~km/s and that of the poloidal component around 30--60~km/s in magnitude, with the toroidal in the negative $\whbs{\phi}$ direction and the poloidal in the positive $\whbs{\theta}$ direction.  The radial component grows from around 20~km/s to around 50~km/s in magnitude, in the positive $\whbs{r}$ direction.
Similarly to the octahedral case, the tetrahedral measurements for the plasma startup and rampdown phases tend to exhibit faster changes than do those for the flattop.

\noindent

\section{Discussion}
\label{sec:discussion}

Though sharing the same overall order of magnitude, lending confidence that our general matrix method is sound in principle, our example flow velocity measurement results from the octahedral and tetrahedral Mach probes are at the same time markedly different from each other, which is one motivation to consider them as initial and preliminary.
Thus, they also exhibit different comparisons to previous measurements of spontaneous equilibrium flow velocities in MST
\cite{fontana_1999, miller_2011}.
Miller
\cite{miller_2011} used a 2D Mach probe for an insertion depth scan of toroidal and poloidal flows, finding flow speeds and directions roughly similar to those we measured with the octahedral probe though not precisely matching.
Fontana
\cite{fontana_1999}
used a rotatable IDS probe
\cite{fiksel_1998}
to measure all three components of equilibrium flow across multiple shots, finding in particular that the measured flow directions were consistent with E-cross-B.
Without going into much detail, we note that our octahedral results tend to agree more closely than do our tetrahedral results in magnitude with the previous results.
By contrast, while our tetrahedral results are roughly consistent with E-cross-B in overall velocity direction, the measured poloidal and radial magnitudes are much larger than expected, being in the neighborhood of $M=1$.

It is conceivable that for some reason, the tetrahedral probe happened to suffer more than did the octahedral probe from conditioning effects such as intermittent plasma arcing.
Fast electrons, a known issue in MST RFP plasmas
\cite{stoneking_1994},
might have significantly affected the ion saturation current measurements.
Related to this, perhaps one key difference is the ratio of Mach tip bias voltage to $\Te$, which was substantially larger for the octahedral case than for the tetrahedral.
It might be expected that a larger ratio could improve the reliability of the ion saturation current measurements in the presence of fast electrons.
We also continue to double-check things like signal channel identities and probe rotation conventions, which clearly could have a major effect on the results if there were a problem.

In summary, we have developed a novel matrix method for analyzing ion saturation current data from general 3D unmagnetized Mach probe arrays including regular octahedral and tetrahedral arrays.
While analysis of octahedral probe data is a straightforward extension of the 1D upstream-downstream Mach probe methodology, the tetrahedral analysis is specifically enabled by this new method, as it does not involve any pairs of directly opposing tips.
Using tetrahedral Mach probes can provide an advantage in compactness compared to octahedral probes.
We have calculated velocity measurement uncertainties within this framework and estimate that the combined uncertainty due to finite machine precision and ion saturation current measurement uncertainty is of order several percent, which is likely smaller than systematic uncertainties involved in Mach probe modeling
\cite{chung_2012}
and experimental uncertainties relating to probe conditioning.
We have applied the matrix method to experimental data from octahedral
\cite{sellner_2024}
and tetrahedral
\cite{rouda_2025}
Mach probes in RFP plasmas in MST to produce initial measurements of the local 3D plasma flow velocity in both cases.
We find that the basic orders of magnitude of the velocity components are reasonable, though the clear disagreements that do exist motivate further attention to these data sets.

Future work is planned to use our tetrahedral probe data set for a study of the behavior of cross helicity density $\bs{v}\cd\bs{B}$, a main ingredient of canonical helicity
\cite{you_2012, you_2014, von_der_linden_2018}.
We also note that, as mentioned in Sec.~\ref{sec:experiments}, the tetrahedral probe we used here
\cite{rouda_2025}
was part of an array of multiple tetrahedral probes designed to measure the fluid vorticity $\curl\bs{v}$, another ingredient of the canonical helicity.
This purpose was not achieved due to hardware issues with the other probes in the array, highlighting the challenges of minimizing the number of required measurements and probe size.
It remains promising that 3D Mach probes of the types analyzed and tested here could provide local plasma flow measurements for this and other future laboratory investigations.

\begin{acknowledgments}
This material is based upon work supported by the U.S. Department of Energy, Office of Science, Office of Fusion Energy Sciences, under Award Number DE-SC0018266.
\end{acknowledgments}

%

\end{document}